# Tuning photovoltaic response in $Bi_2FeCrO_6$ films by ferroelectric poling

A. Quattropani,[a] A. S. Makhort,[b] M. V. Rastei,[b] G. Versini,[b] G. Schmerber,[b] S. Barre,[b] A. Dinia,[b] A. Slaoui,[a] J.-L. Rehspringer,[b] T. Fix,[a] S. Colis[b] and B. Kundys*[b]

Ferroelectric materials are interesting candidates for future photovoltaic applications due to their potential to overcome the fundamental limits of conventional single bandgap semiconductor-based solar cells. Although a more efficient charge separation and above bandgap photovoltages are advantageous in these materials, tailoring their photovoltaic response using ferroelectric functionalities remains puzzling. Here we address this issue by reporting a clear hysteretic character of the photovoltaic effect as a function of electric field and its dependence on the poling history. Furthermore, we obtain insight into light induced nonequilibrium charge carrier dynamics in $Bi_2FeCrO_6$ films involving not only charge generation, but also recombination processes. At the ferroelectric remanence, light is able to electrically depolarize the films with remanent and transient effects as evidenced by electrical and piezoresponse force microscopy (PFM) measurements. The hysteretic nature of the photovoltaic response and its nonlinear character at larger light intensities can be used to optimize the photovoltaic performance of future ferroelectric-based solar cells.

## Introduction

The non-centrosymmetric structure of large bandgap polar materials induces an internal electric field comparable to that existing in the p–n junction region of semiconductor based solar cells.[1,2] Subsequently, electrically polar materials with photovoltaic (PV) properties have gained renewed attention with regard to photovoltaics[3–10] and other attractive multi-functionalities.[11–21] Although the photovoltaic effect in non-centrosymmetric crystals has been known for long,[22] this field has attracted increasing attention following the discovery of photovoltaic effects in the multiferroic $BiFeO_3$(BFO).[23,24] Several ferroelectric (FE) materials in the thin film form such as $BiMnO_3$ (BMO),[25] La and Ni-doped $Pb(Zr,Ti)O_3$ (PLZT),[26,27] $BaTiO_3$ (BTO)[28] and bulk $[KNbO_3]_{1-x}[BaNi_{1/2}Nb_{1/2}O_{3-\delta}]$ (KBNNO)[5] have been studied over the last few years, but none of them were able to generate a remarkable efficiency, mainly because of their bandgap being larger than 2.5 eV. In contrast, the multiferroic BFO and related materials[3,4,29,30] showing moderate energy bandgaps manifest interesting PV properties. Moreover, the bandgap of BFO can be varied by doping and by modulating the preparation conditions. Indeed, Cr doping of BFO results in a double perovskite $Bi_2FeCrO_6$ (BFCO) structure with different amounts of Fe–Cr cationic disorder, which allows bandgap engineering down to 1.9 eV. Nechache *et al.*[31] claimed that in perfectly ordered BFCO the bandgap can be even lower (1.4 eV) and succeeded to fabricate BFCO based solar cells with efficiencies of 3.3% and 8.1% using a single layer and a tandem-like configuration, respectively. These efficiencies are expected to improve further because, at least theoretically, solar cells integrating ferroelectric materials could exceed the Shockley–Queisser limit.[32] Moreover, the increase of efficiency in FE solar cells also requires a deep insight into their dependence on the FE state, and the charge generation–evolution mechanisms. Here we investigate these issues by using careful photovoltaic response measurements *versus* electric field and demonstrate how the PV effect can be tuned by varying the applied voltage and related poling history. We also study the light intensity dependence of the PV effect aiming at the fundamental understanding of the ferroelectric based photovoltaic performance.

## Results and discussion

In order to perform the electrical measurements, a 100 nm thick transparent ITO layer was sputtered at room temperature on the BFCO film, providing both good electrical conductivity and high optical transparency (Fig. 1, left). The 1 mm² area of the electrode was then brought into contact with a copper wire using a conductive epoxy (Fig. 1, right). Fig. 2 shows the $\theta$–$2\theta$ X-ray diffractogram recorded for $Bi_2FeCrO_6$ deposited on the Nb:STO substrate (NSTO). In addition to the 00h peaks of the NSTO substrate, we observe the 001 peaks of the BFCO films. The position of the peaks suggests an epitaxial growth of the pseudo-cubic BFCO phase with the 001 axis parallel to the growth direction. No secondary phases could be observed in the resolution limit of the X-ray diffraction technique. The out-of-plane (OP) lattice parameter $c$ of the BFCO layer was determined to be 0.3965 nm, close to the lattice parameter of bulk $Bi_2FeCrO_6$ (0.3930 nm). The epitaxial growth of our films was unambiguously demonstrated by $\Phi$ scan measurements.[33]

[a]*Université de Strasbourg, CNRS, Laboratoire ICube, UMR 7357, F-67000 Strasbourg, France*
[b]*Université de Strasbourg, CNRS, Institut de Physique et Chimie des Matériaux de Strasbourg, UMR 7504, 23 rue du Loess, F-67000 Strasbourg, France.*
*E-mail: kundys[AT]ipcms.unistra.fr*

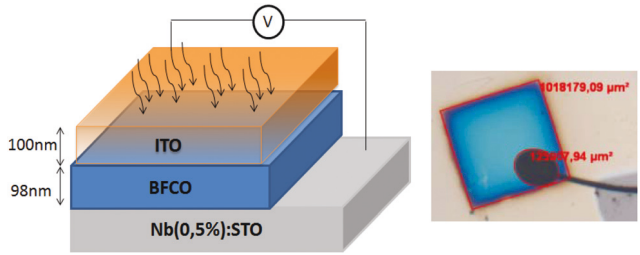

**Fig. 1** Schematic representation of the device used for the experiment (left) and a microscopy image (right) of the electrical contact indicating the active sample area.

Given the low doping level (<1% of Nb) of NSTO substrates, the same epitaxial quality is obtained when BFCO films are grown on STO and NSTO. The optical properties of the as-grown BFCO films were also investigated by UV-Vis-NIR spectrophotometry measurements. Fig. 3a shows the normalized absorption spectra recorded on the STO substrate and on a BFCO film grown on STO. The film strongly absorbs in a large range from 200 to 800 nm (Fig. 3a). Fig. 3b illustrates the corresponding Tauc plot,[34] where the optical absorption coefficient ($\alpha$) relates to bandgap ($E_g$) via Planck's constant ($h$) and the frequency of the incident photon ($\nu$) as[35–37]: $\alpha = (h\nu - E_g)^{1/2}$.

Assuming a direct allowed bandgap for the BFCO film, the bandgap of 1.66 ± 0.04 eV is evaluated. This value is consistent with our previous work.[33] Note however that a shoulder is also observed on the Tauc plot below this energy. Nechache et al.[31] attributed this shoulder to the absorption edge (1.4 eV) of perfectly ordered BFCO (i.e. showing a perfect alternation between Fe and Cr cations) while the second absorption edge (1.66 eV) was attributed to BFCO presenting Fe–Cr disorder.

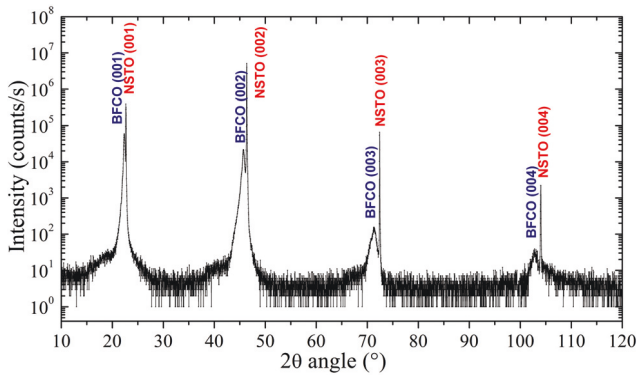

**Fig. 2** XRD $\theta$–$2\theta$ pattern of a BFCO thin film grown on the NSTO (001) substrate.

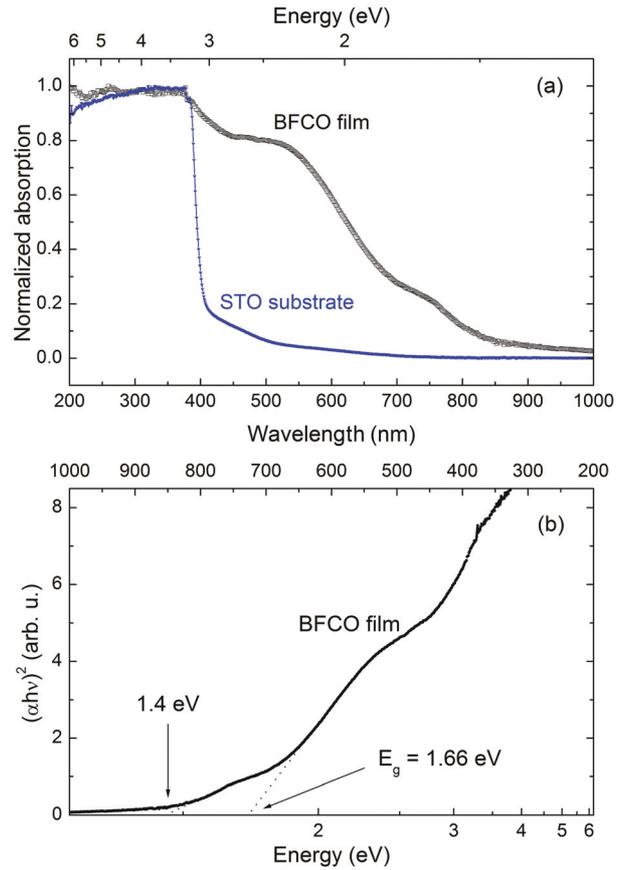

**Fig. 3** Absorption spectrum of BFCO deposited on STO (a) and the corresponding Tauc plot (b) indicating two absorption edges at 1.4 and 1.66 eV.

The electrical measurements were first performed in darkness by sweeping voltage between ±8 V to test ferroelectricity (Fig. 4). At +6 V, a clear FE peak originating from the polarization reversal is observed (Fig. 4 (inset)) even though the curve shows a large level of leakage current. Although the expected peak is not visible for negative voltages due to the overwhelming ohmic conductivity, the FE coercive field of ~64 MV m⁻¹ extracted from the volt–ampere characteristic is in a good agreement with the previous study[38] and the polarization switching can also be evidenced by PFM measurements (Fig. 7). In order to study the effect of light on the BFCO sample, a current–voltage measurement has been performed in the dark and under 365 nm wavelength excitation light emitted by a light emitting diode (LED) with 30 nm spectral linewidth. Because the photovoltaic effect in ferroelectrics can be poling dependent,[39–41] the sample was previously depolarized in darkness using an oscillating damped voltage application pro-cedure. After that, the sample was exposed to 365 nm and 153 mW cm$^{-2}$ of radiant power, and the current was recorded in the depolarized state during voltage sweep starting from zero to +8 V. Afterwards, the voltage was swept from +8 V to −8 V (marked as positive to negative (P → N) sweep) followed by the reverse negative to positive (N → P) sweep (from −8 V to +8 V) (Fig. 5). The FE-history dependence of the photovoltaic effect becomes clearly evident with an important short-circuit photocurrent ($I_{sc}$) evolution. $I_{sc}$ can be effectively tuned between 0.9 and 2.2 µA depending on the poling sweep. It also becomes evident that the presence of FE domains (depolarized state) clearly affects the photovoltaic effect. It has to be noted that the role of domain walls in the PV effect is a long debated issue.[29,42–45] In that respect, our results clearly demonstrate that the polydomain (depolarized) state can indeed be advantageous for the PV effect under specific conditions (if com-pared with the random FE ground state or here with the state under the PN electric field sweep). However, a proper electrical pooling (or here NP sweep) can show even better photovoltaic response than that of the depolarized state. The insight into the PV hysteresis is therefore clearly needed to optimize the performance. A similar hysteretic behavior is also detected for the open circuit voltage ($V_{oc}$) with an average value of 0.54 V. To obtain more insight into the photovoltaic memory effect, the open circuit voltage was measured as a function of time during the periodic illumination of the sample (Fig. 6). Prior to measurements, the sample was polarized positively by sweeping voltage from −8 V → +8 V → 0 V. The sample was then first illuminated for 4 minutes and then the light was switched off for 8 minutes. Subsequently, the LED was switched ON and OFF three more times (Fig. 6).

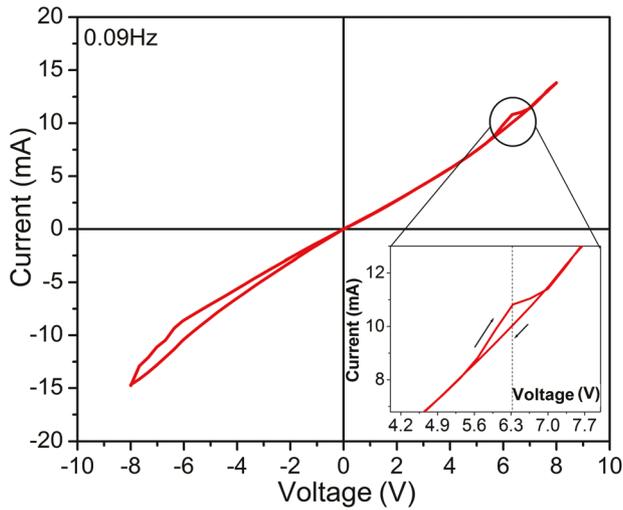

Fig. 4 *I–V* loop of the BFCO film recorded in darkness.

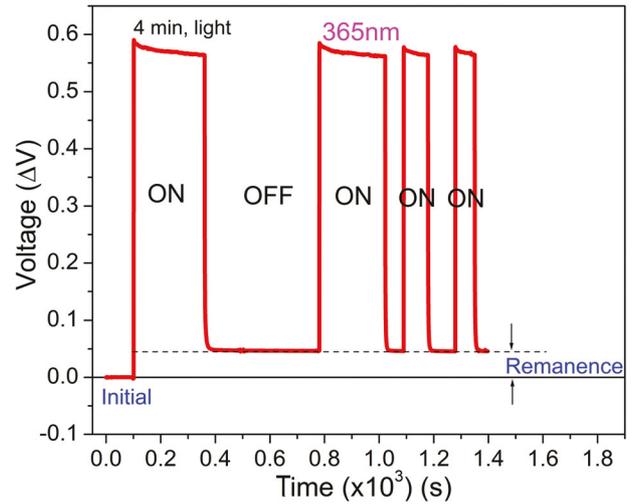

Fig. 6 Transient and remanent photovoltaic effect for the BFCO sample.

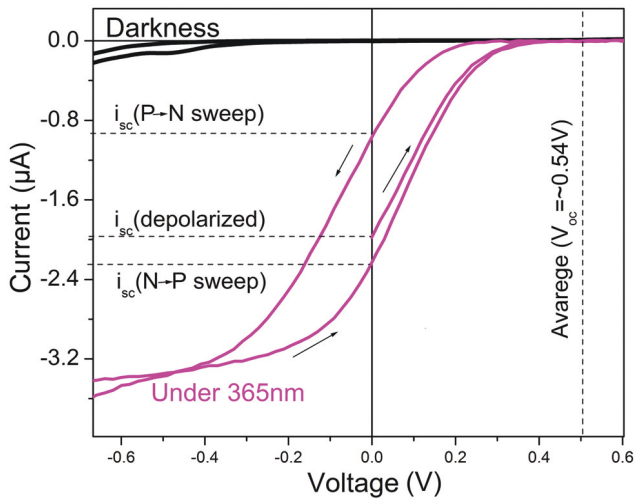

Fig. 5 Current–voltage measurements on a BFCO sample, in darkness and under 365 nm light irradiation.

The curve features show that at FE saturation the photovoltaic response is in fact a combination of the optically reversible (transient) and irreversible (remanent) effects. With the sample initially polarized positively, the first pulse of light induces a jump in photovoltage reaching a value of 0.56 V. This is followed by an interesting behavior since upon switching off the light, the voltage value did not drop down to the initial starting point, but to a value of 50 mV giving rise to a remanent voltage value. All the subsequent light pulses show the same reversible effect. This behavior can be attributed to the light induced change in the polarization state featuring both reversible and irreversible components as was recently shown by Makhort *et al*.[41] In an attempt to elucidate the observed phenomena, PFM measurements on the film surface free of the ITO electrode layer have been carried out.

The ferroelectric nature of the film is confirmed by the presence of the electrically switchable polarization states (Fig. 7). Poling images were obtained by applying +8 V (large square) followed by −8 V (small square) (Fig. 7). The electrically written FE domains clearly lose their contrasts after light exposure for 4 min (which is the same time used for the first pulse in Fig. 6). In agreement with Fig. 6, the further light exposure does not modify the PFM images and the initial contrast can only be recovered electrically. The observed behavior can be explained as follows. For a sample initially polarized positively, the illumination generates carriers which distribute along the previously forced polarization direction. The generated photo-carriers reduce the surface's charges, and consequently change the internal electric field of the material depolarizing the sample. Upon turning off the light, the carrier trapping makes this decrease partly persisting (remanent effect), leaving the sample in a different polarization state. All the subsequent pulses of illumination reveal only a reversible effect; these pulses are in fact of the same power and the corresponding trapping centers are already occupied.

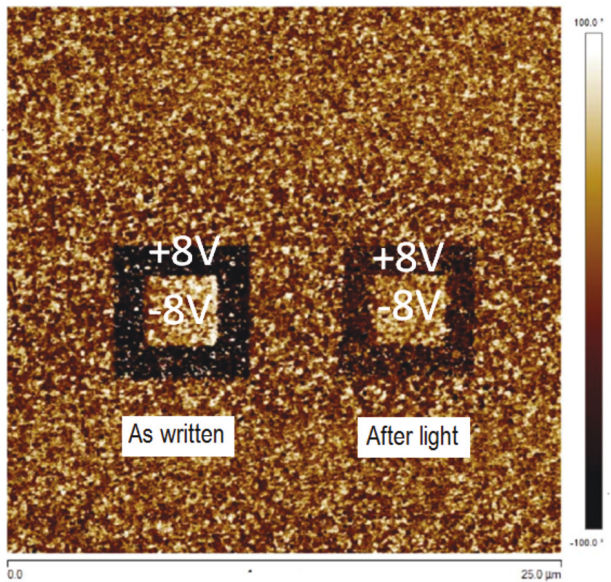

Fig. 7  PFM poling images before and after exposure to 365 nm light.

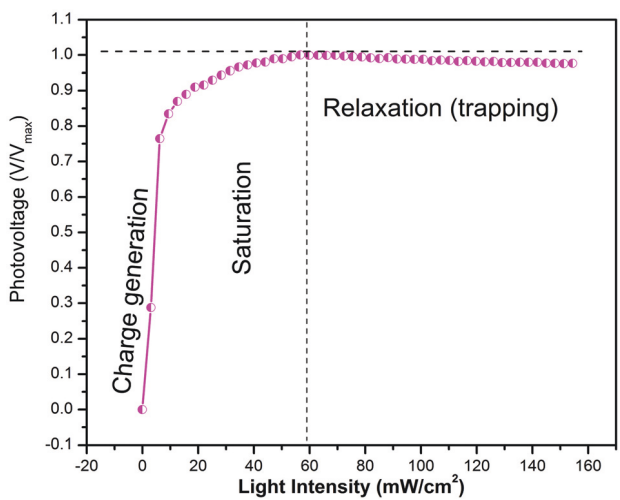

Fig. 8  Relative change in voltage induced by 365 nm illumination plotted versus light intensity.

The hypothesis of carriers' trapping effect can be confirmed by measuring the photovoltage versus the light intensity (Fig. 8). The curve exhibits the nonlinear behavior with three different regions: a fast charge generation, an intermediate saturation and a slow relaxation. The latter can be possibly attributed to the charge trapping processes involved in the remanent effect as shown in Fig. 6. It is noteworthy that the possible effect of the light-induced increase of temperature in the sample during illumination can be discarded. The measured temperature change using a thermal camera exhibited only a 1 K temperature increase compared to the unexcited sample. The sample was then heated in a cryostat and it was found that such heating induces a negligible contribution to the sample pyrocurrent and voltage change compared to the light induced effects.

## Experimental

Epitaxial $Bi_2FeCrO_6$ (BFCO) films studied in this work were grown on $SrTiO_3$ (STO) or Nb-doped (0.5%):$SrTiO_3$ (001) (NSTO) by Pulsed Laser Deposition (PLD) using a KrF (248 nm) laser and a home-made target. The laser fluence was about 2 J cm$^{-2}$ and the repetition rate was 2 Hz. The deposition was carried out at 750 °C under an oxygen pressure of $10^{-2}$ mbar. The as-grown film was cooled down at 5 °C min$^{-1}$ rate under the same atmosphere. A film thickness of 98 nm was measured using X-Ray Reflectivity (XRR). The crystalline structure was investigated by X-ray diffraction (XRD) using a SmartLab Rigaku diffractometer equipped with a Cu source and a Ge (220 × 2) crystal delivering the monochromatic CuK$_{α1}$ radiation (0.154056 nm). The high angle measurements allowed identifying the crystalline structure, calculating the size of the crystallites, and checking the phase purity and epitaxial quality of the films. The optical measurements of the films were investigated in the 200–1200 nm range using a Perkin-Elmer Lambda 950 spectrophotometer working in the ultraviolet-visible-near infrared (UV-Vis-NIR) range. The ferroelectric properties were investigated by the Piezoresponse Force Microscopy (PFM) technique using a Bruker Icon QNM microscope. The tip is a 0.01–0.025 Ohm cm antimony (n) doped Si covered with conductive diamond. A 4 V AC voltage was applied to the tip at 20 kHz. In this configuration, the current–voltage measurements were performed using a precision LCR Meter (Agilent E4980A).

## Conclusions

In this work, the successful preparation of epitaxial BFCO films on Nb:STO substrates allowed the observation of ferroelectric and photovoltaic effects demonstrating a clear link between the two. Most importantly, the poling history dependence of the photovoltaic effect, clearly reported here for the first time, can be regarded as a key basic finding opening a way to electric field tuning of the photovoltaic performance and shedding light on the long-debated issue about the role of domain walls in the PV effect. Moreover, the photoelectric characterization reveals a nonlinear character of the photovoltaic response *versus* light intensity involving change generation, saturation and recombination processes. The latter effects can be used for all-optical information storage exploiting light induced remanent changes in polarization. More generally, the results reported here are not often encountered, thereby stimulating a new path of research towards electrically optimized ferroelectric-based solar cells.

## Author contribution

A.Q. as a PhD student participated in the thin film preparation and photovoltaic–ferroelectric measurements under the supervision of A.S., A.D., T.F., and B.K., respectively. M.V.R. and A.S.M. performed AFM experiments under light. G.V. and S.B. conducted the film deposition. G.S. performed X-ray and optical absorption measurements. J.-L. R. participated in discussions and planning. A.S., A.D., T.F., S.C. and B.K. managed the discussions and analyses and co-wrote the paper.

## Conflicts of interest

There are no conflicts to declare.

## Acknowledgements

This work was carried out under the framework of the FERROPV project supported by the French Agence Nationale de la Recherche (ANR) under the reference ANR-16-CE050002-01. The partial support of the ANR (ref. ANR-11-LABX-0058-NIE within the Investissement d'Avenir program ANR-10-IDEX-0002-02) is also acknowledged.

## Notes and references